\documentclass[aps,twocolumn,floatfix,preprintnumbers,
               tightenlines,showpacs,showkeys,notitlepage,nofootinbib]{revtex4-2}

\usepackage[colorlinks=true,citecolor=blue,linkcolor=blue]{hyperref}
\usepackage[normalem]{ulem}
\usepackage{amsmath,amssymb}
\usepackage{mathtools}
\usepackage{verbatim}
\usepackage{titlesec}               
\usepackage{graphicx}               
\usepackage{wasysym}
\usepackage{siunitx}
\usepackage{cleveref}
\usepackage{subfigure} 
\usepackage{xspace}                 
 
\allowdisplaybreaks

\makeatletter

\renewcommand{\p@subsection}{}
\makeatother

\titleformat*{\section}{\centering\bfseries\uppercase}
\titlelabel{\thetitle\quad}
\titleformat*{\paragraph}{\bfseries}
\titlespacing*{\paragraph}{0pt}{3.25ex plus 1ex minus .2ex}{1em}

\begin{document}

\title{Resonant Spin-Flavor Precession of Sterile Neutrinos}

\author{Edward Wang}
\email{edward.wang@tum.de}
\affiliation{Physik Department T70, Technische Universität München, James-Franck-Straße, 85748 Garching, Germany}

\preprint{TUM-HEP-1484/23}

\begin{abstract}
\noindent
We analyze the impact of resonant conversions mediated by non-vanishing magnetic moments between active neutrinos and a heavy sterile neutrino on the supernova neutrino flux. We present the level-crossing scheme for such a scenario and derive the neutrino fluxes after conversion, paying particular attention to the order in which the resonances occur. We then compute the expected event rates from the neutronization burst of a future supernova at DUNE and Hyper-Kamiokande to derive new constraints on the neutrino magnetic moment. With this, we find a sensitivity down to a few $\SI{e-15}{\mu_B}$ for a sterile neutrino in the $O(\SI{}{eV})$ mass range.
\end{abstract}

\maketitle

\section{Introduction}
\label{sec:intro}

Shortly after the neutrino was first hypothesized by Pauli in 1930 \cite{Pauli:1930pc}, the first studies on the magnetic moment of the neutrino were already conducted, with a first experimental constraint established already in 1935 \cite{Nahmias:1935yih}. In recent years, the possibility of probing heavy sterile neutrinos through the magnetic moment has been intensely explored, with measurable effects ranging from terrestrial experiments, astrophysical phenomena, and cosmology \cite{Magill:2018jla,Brdar:2020quo,Brdar:2023tmi}.

In the present work, we revisit the resonant magnetic conversion, often called resonant spin-flavor precession (RSFP), of active neutrinos into sterile neutrinos inside a supernova, first investigated in ref. \cite{Nunokawa:1998vh}. The mechanism is similar to magnetic transitions between active neutrinos \cite{Akhmedov:1987nc,Lim:1987tk,Akhmedov:1988uk,Ahriche:2003wt,Jana:2022tsa}, and Mikheev-Smirnov-Wolfenstein (MSW) conversions between active and sterile neutrinos \cite{Kainulainen:1990bn,Raffelt:1992bs,Shi:1993ee,Arguelles:2016uwb}, which have been studied extensively in the literature. These transitions can either convert part of the active neutrino flux into sterile neutrinos or, through successive transitions, change the flavors of the neutrinos as they propagate inside a supernova.

The structure of this paper is the following: in \cref{sec:mm} we formulate the basic ingredients for describing the magnetic conversion of active and sterile neutrinos. In \cref{sec:sn} we apply this formalism to a supernova environment and detail how the transitions at resonances affect the final fluxes. We then compare the expected rates of events with and without a magnetic moment at the future neutrino detectors DUNE and Hyper-Kamiokande in \cref{sec:exp}. We summarize and discuss our results in \cref{sec:cn}.

\section{Neutrino Magnetic Moments}
\label{sec:mm}

We consider a scenario where the Standard Model is extended by a right-handed singlet fermion $N$ of mass $M_N$, often called sterile neutrino or heavy neutral lepton, which couples to active neutrinos via a magnetic moment. These magnetic moment interactions are described by the effective operator
\begin{align}
    \mathcal{L} \supset \frac{\hat{\mu}_\nu^{\alpha N}}{2} F_{\mu \nu} \bar{\nu}_\alpha \sigma^{\mu \nu} N + \text{h.c.},
    \label{eq:Lagrangian}
\end{align}
where $\nu^\alpha$ is the left-handed neutrino field of flavour $\alpha$, and $F_{\mu \nu}$ is the electromagnetic field strength tensor. In this work, we make no distinction between $N$ being a Majorana or a Dirac fermion. In principle, an important difference between the Dirac and Majorana cases is that, if $N$ were Majorana, one could have transitions $\nu_L \to N_R \to N_R^c \to \nu_L^c$, however the $N_R \to N_R^c$ oscillation is suppressed by a factor of $M_N^2/p^2$ \cite{Bahcall:1978jn,Schechter:1980gk,Li:1981um,Bernabeu:1982vi,Kimura:2021juc}, where $p$ is the momentum of $N$. Given that we only consider $M_N$ up to $\sim \SI{10}{keV}$, and typical energies of supernova neutrinos are in the MeV range, this effect is negligible for the present work. The neutrino flavour eigenstate fields $\nu_L^\alpha$ are related to their mass eigenstates, $\nu^i$ as to
\begin{align}
    \nu_\alpha = U_{\alpha i} \nu_i \,,
\end{align}
where
\begin{align}
    U = \begin{pmatrix}
          1 &         &        \\
            &  c_{23} & s_{23} \\
            & -s_{23} & c_{23}
        \end{pmatrix}\!\!\!
        \begin{pmatrix}
          c_{13}              &   & s_{13} e^{i\delta} \\
                              & 1 &                    \\
          -s_{13} e^{i\delta} &   & c_{13}
        \end{pmatrix}\!\!\!
        \begin{pmatrix}
          c_{12} & s_{12} &   \\
         -s_{12} & c_{12} &   \\
                 &        & 1
        \end{pmatrix} \notag
\end{align}
is the Pontecorvo–Maki–Nakagawa–Sakata (PMNS) matrix, with $s_{ij} \equiv \sin\theta_{ij}$, $c_{ij} \equiv \cos\theta_{ij}$, and the CP-violating phase $\delta$.

The flavor evolution of a neutrino is governed by the Schr\"odinger equation
\begin{align}
    i \frac{\text{d}}{\text{d}t} \psi(t) = \hat{H} \psi(t) \,,
    \label{eq:schroedinger}
\end{align}
where $\psi$ is a unit vector in flavor space whose components describe the admixture of each flavor to the neutrino state. In the flavor basis, $\psi$ is given by $\psi = (\nu_e, \nu_\mu, \nu_\tau, N)$ and we can write the Hamiltonian in block diagonal form as
\begin{align}
    \hat{H} &= \frac{1}{2p}
               \begin{pmatrix}
                   U & \\
                   & 1
               \end{pmatrix}
               \begin{pmatrix}
                   \hat{M}_\nu^2 & \\
                   & M_N^2
               \end{pmatrix}
               \begin{pmatrix}
                   U^\dag & \\
                   & 1
               \end{pmatrix}
                                    \notag\\[0.2cm]
            &+              
                \begin{pmatrix}
                   \hat{H}_\text{MSW} & \\
                    & 0
               \end{pmatrix}
               + \frac{B_\perp}{2}
               \begin{pmatrix}
                   0 & \hat{\mu}_\nu \\
                   \hat{\mu}_\nu^\dagger & 0
               \end{pmatrix} \,.
    \label{eq:H}
\end{align}
with the neutrino momentum $p$, the diagonal mass matrix
\begin{equation}
    \hat{M}_\nu = \text{diag} ( m_{\nu_1}, m_{\nu_2}, m_{\nu_3} ) \,,  \label{eq:M-nu}
\end{equation}
the MSW potential
\begin{equation}
	\hat{H}_\text{MSW} =
	\begin{pmatrix}
		V_{\nu_e} & & \\
		& V_{\nu_\mu} & \\
		& & V_{\nu_\tau}
	\end{pmatrix} \,,
\end{equation}
and the magnetic moment vector
\begin{align}
    \hat{\mu}_\nu = \begin{pmatrix}
                        \mu_\nu^{eN}  \\
                        \mu_\nu^{\mu N}  \\
                        \mu_\nu^{\tau N}
                    \end{pmatrix} \,.
    \label{eq:mu}
\end{align}

In principle, the perpendicular magnetic field $B_\perp$ also contains a geometric phase $\phi$ defined by $B_x + i B_y = B_\perp e^{i \phi}$, assuming $z$ is the propagation axis of the neutrinos. Variations of this geometric phase can alter the resonant behavior or give rise to new resonances \cite{Smirnov:1991ia,Akhmedov:1991vj,Balantekin:1993ys}. The primary impact of this in our scenario would be a shift by $\dot{\phi}$ in the resonance condition \cite{Jana:2023ufy}, which in our case can be compensated by a shift in $M_N$ (s. \cref{sec:sn}). Additionally, if $\ddot{\phi}$ is large, it can also decrease the adiabaticity parameter, leading to a weaker sensitivity. In the absence of precise estimates of both $\dot{\phi}$ and $\ddot{\phi}$, we don't consider this effect in the present work.

We also note that the large mass gap between the active and the sterile neutrinos suppresses spin-flavor oscillations in vacuum, unlike in the Dirac case \cite{Kurashvili:2017zab,Lichkunov:2022mjf,Alok:2022pdn,Kopp:2022cug,Alok:2023sfr}. We further neglect mass mixing between the active neutrinos and $N$.

Assuming equal number densities of electrons and protons and defining the electron number fraction $Y_e = n_e/(n_e + n_p)$, the MSW potentials can be written as
\begin{align}
	V_{\nu_e} =& \frac{\sqrt{2}}{2} G_F \frac{\rho}{m_N} (3 \, Y_e - 1), \label{eq:MSW_potential1} \\
	V_{\nu_{\mu, \tau}} =& \frac{\sqrt{2}}{2} G_F \frac{\rho}{m_N} (Y_e - 1),
    \label{eq:MSW_potential2}
\end{align}
where $G_F$ is the Fermi constant, $\rho$ is the matter density and $m_N$ the nucleon mass.

\section{Neutrino Conversion inside the Supernova}
\label{sec:sn}

\begin{figure}
	\includegraphics[width=\columnwidth]{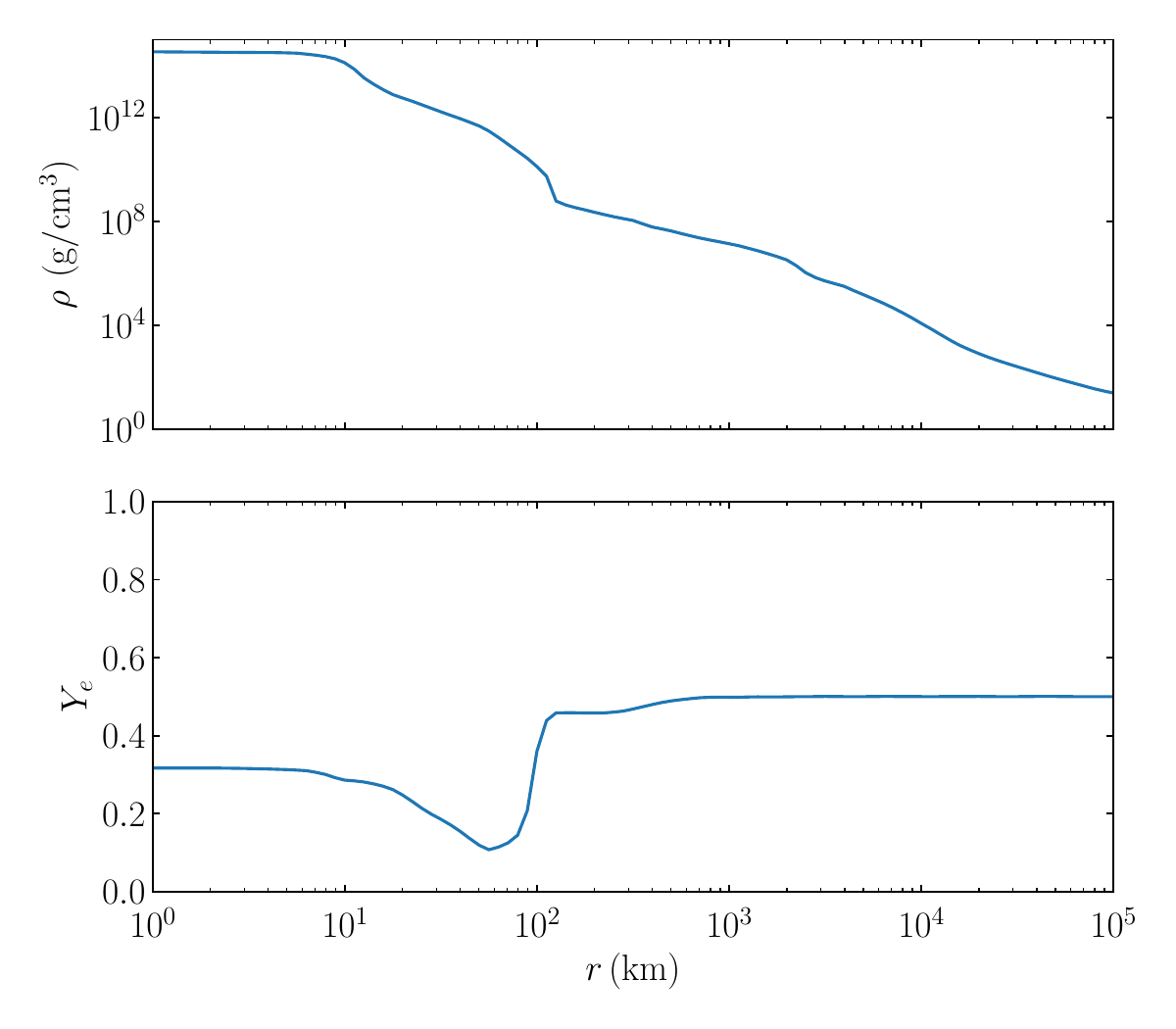}
	\caption{Density and electron fraction profiles for a $10.8 M_{\astrosun}$ progenitor supernova at $t = \SI{30}{ms}$ after core bounce, taken from ref. \cite{Fischer:2009af}.}
	\label{fig:profiles}
\end{figure}

In the current work, we analyze the neutrino flux during the neutronization burst, in which a large number of neutrinos are emitted in a short period of time. The choice of looking at the neutronization burst is motivated by the large neutrino flux emitted and by its robust predictions, which are consistent up to $O(10\%)$ among different simulations \cite{Serpico:2011ir,Wallace:2015xma,OConnor:2018sti,Kachelriess:2004ds}; in this work we use the initial flux from ref. \cite{Hudepohl:2009tyy} for a supernova with a $8.8 \, M_{\astrosun}$ progenitor. We use the results from ref. \cite{Fischer:2009af} for a $10.8 \, M_{\astrosun}$ progenitor at $\SI{30}{ms}$ after bounce for the mass density and electron fraction profiles. In addition to this, we model the magnetic field within the supernova as a dipole $B(r) = 2 B_0 r_0^3/(r^3 + r_0^3)$ \cite{Suwa:2007nq}, where $B_0$ is the magnetic field at the iron surface, and $r_0 = \SI{1671}{km}$ is the radius of the iron core. Throughout this work we set $B_0 = \SI{e12}{G}.$

\begin{figure*}
    \begin{tabular}{c@{}c}
        \hspace*{-0.7cm}
        \includegraphics[width=1.1\columnwidth]{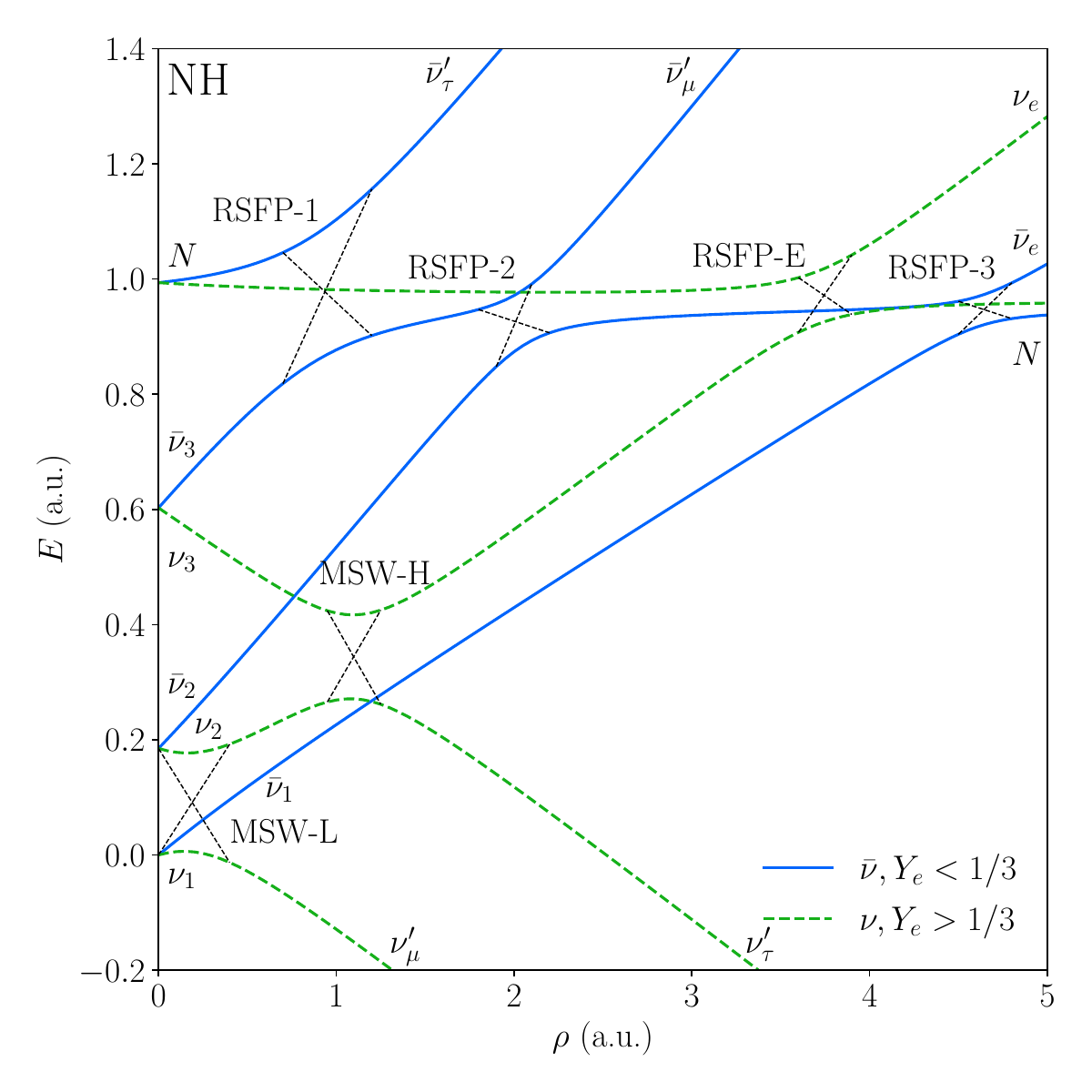}
        &
        \hspace*{-0.3cm}
        \includegraphics[width=1.1\columnwidth]{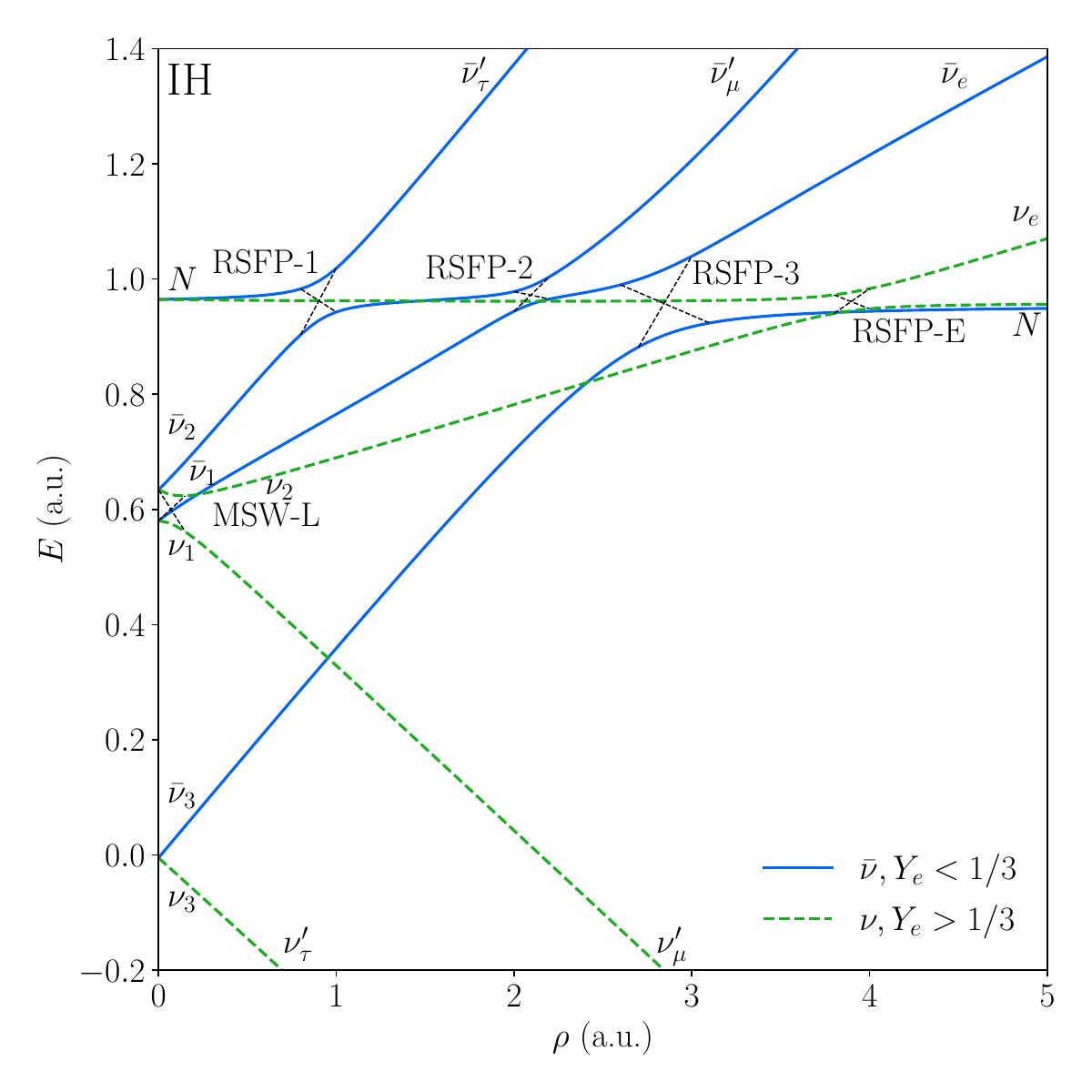}
    \end{tabular}
	\caption{Level-crossing scheme for normal (left) and inverted (right) hierarchy. For neutrinos we set $Y_e > 1/3$, while for antineutrinos we choose $Y_e < 1/3$, so that both RSFP-E and RSFP-3 are present (s. discussion of resonance conditions after \cref{eq:res_cond_E}).}
	\label{fig:crossing}
\end{figure*}

There has been significant interest in magnetic fields of core-collapse supernovae in recent years, both due to their potential impact on the explosion mechanism and to explain the formation of highly magnetized neutron stars, so-called magnetars. In particular, recent magnetohydrodynamic simulations have found that initial magnetic fields of the order of $\SI{e12}{G}$ can be amplified by up to three orders of magnitude via magnetorotational instabilities \cite{Rembiasz_2016} as well as standing accretion shock instabilities \cite{Varma:2022piv} a few milliseconds after bounce.

Since large potential differences damp transitions mediated by the magnetic moment, we only need to consider transitions at resonances, where energy differences are small. We find that, apart from the usual MSW resonances, in this scenario we have additional $\nu \leftrightarrow N$ resonances due to the magnetic moment, which can be computed in a two-flavor approximation.

The resonances occur when the instantaneous eigenvalues of the Hamiltonian are closest to one another; alternatively, for RSFP transitions, they occur where the eigenvalues cross when we turn the magnetic interactions off. In general, the precise resonance point depends on the masses and mixings of the neutrinos as well as the matter potentials. However, since we generally assume $M_N \gg m_\nu$, the resonances occur deep inside the supernova, where the matter potentials dominate the Hamiltonian with respect to the neutrino mass contributions. We can therefore approximate the resonance condition as
\begin{equation}
	V_{\nu_{e, \mu, \tau}} = \pm \frac{M_N^2}{2 p}.
    \label{eq:resonance_condition}
\end{equation}
Inserting the potentials from \cref{eq:MSW_potential1,eq:MSW_potential2} we find four resonances
\begin{align}
    V_{\bar{\nu}_{\mu, \tau}} = \frac{\sqrt{2}}{2} G_F \frac{\rho}{m_N} (1 - Y_e) =& \frac{M_N^2}{2 p}, \qquad \text{(RSFP-1,2)} \label{eq:res_cond_12} \\
    V_{\bar{\nu}_{e}} = \frac{\sqrt{2}}{2} G_F \frac{\rho}{m_N} (1 - 3 Y_e) =& \frac{M_N^2}{2 p}, \qquad \text{(RSFP-3)} \label{eq:res_cond_3} \\
    V_{\nu_e} = \frac{\sqrt{2}}{2} G_F \frac{\rho}{m_N} (3 Y_e - 1) =& \frac{M_N^2}{2 p}. \qquad \text{(RSFP-E)} \label{eq:res_cond_E}
\end{align}
Since $Y_e < 1$, resonances associated to $V_{\nu_{\mu, \tau}}$ are not possible. 
As can be seen in \cref{fig:profiles}, $Y_e$ undergoes a significant change at $r \approx \SI{100}{km}$, going from initially $Y_e < 1/3$ to $Y_e \approx 1/2$. This means that $V_{\nu_e}$ and $V_{\bar{\nu}_e}$ switch sign at this point, so that RSFP-3 only occurs at $r \lesssim \SI{100}{km}$, while RSFP-E occurs at $r \gtrsim \SI{100}{km}$. It should be noted that, while \cref{eq:res_cond_12} seems to indicate that RSFP-1 and 2 happen at the same point, in reality that's not the case, since the active neutrino masses and mixing, while small, break the degeneracy between the two eigenstates of the Hamiltonian. To determine the order of the resonances, we perform a numerical scan of the eigenlevels, with the level-crossing scheme shown in \cref{fig:crossing}. We order the resonances involving anti-neutrinos according to the densities at which they occur as RSFP-1, RSFP-2, and RSFP-3, while we denote the single resonance involving neutrinos RSFP-E.

We define the adiabaticity parameter
\begin{equation}
	\gamma_\text{res} = \left| \frac{8 \mu_\nu^2 B_\perp^2}{d V_\nu/d r} \right|,
	\label{eq:adiabaticity_parameter}
\end{equation}
with which we can estimate the transition probability at a resonance in the Landau-Zener approximation as
\begin{equation}
	P_\text{res} \approx e^{- \frac{\pi}{2} \gamma_\text{res}},
\end{equation}
where $\gamma_\text{res}$ is computed at the resonance point. The impact of the magnetic moment is in mediating adiabatic conversions of the eigenlevels at the resonance, which happens when $\gamma_\text{res}$ is large. In the opposite case, where $\mu_\nu$ goes to zero, so does $\gamma_\text{res}$, and the levels cross exactly at the resonances. In addition to a finite magnetic moment, adiabatic conversion requires that $B_\perp$ be large or $d V_\nu/d r$ be small. As was pointed out in \cite{Jana:2023ufy}, the density decreases as $\sim r^{-3}$, so that $d V_\nu/d r \sim r^{-4}$, while $B_\perp \sim r^{-3}$ for $r > r_0$. Therefore, $\gamma_\text{res} \sim r^{-2}$, and adiabatic conversion is disfavored at large $r$. At the same time, for $r < r_0$, $B_\perp$ remains approximately constant, while $V_\nu$, and therefore $d V_\nu/d r$, continues to increase as $r$ decreases, so that adiabatic conversion is again disfavored. We therefore expect the largest effect to occur for resonances happening at $r \sim r_0$.

\begin{figure*}
    \centering
    \begin{tabular}{c@{}c}
        \hspace*{-0.7cm}
        \includegraphics[width=1.1\columnwidth]{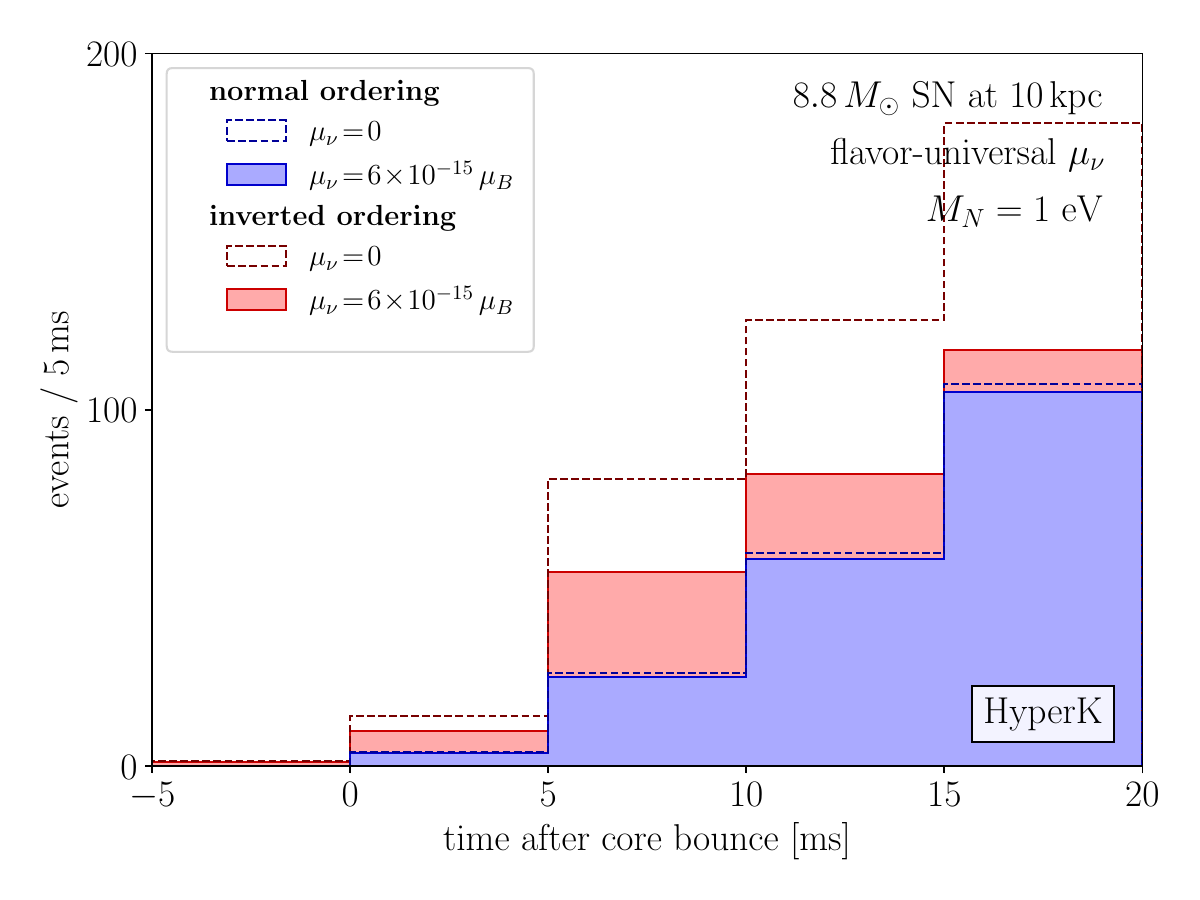}
        &
        \hspace*{-0.2cm}
        \includegraphics[width=1.1\columnwidth]{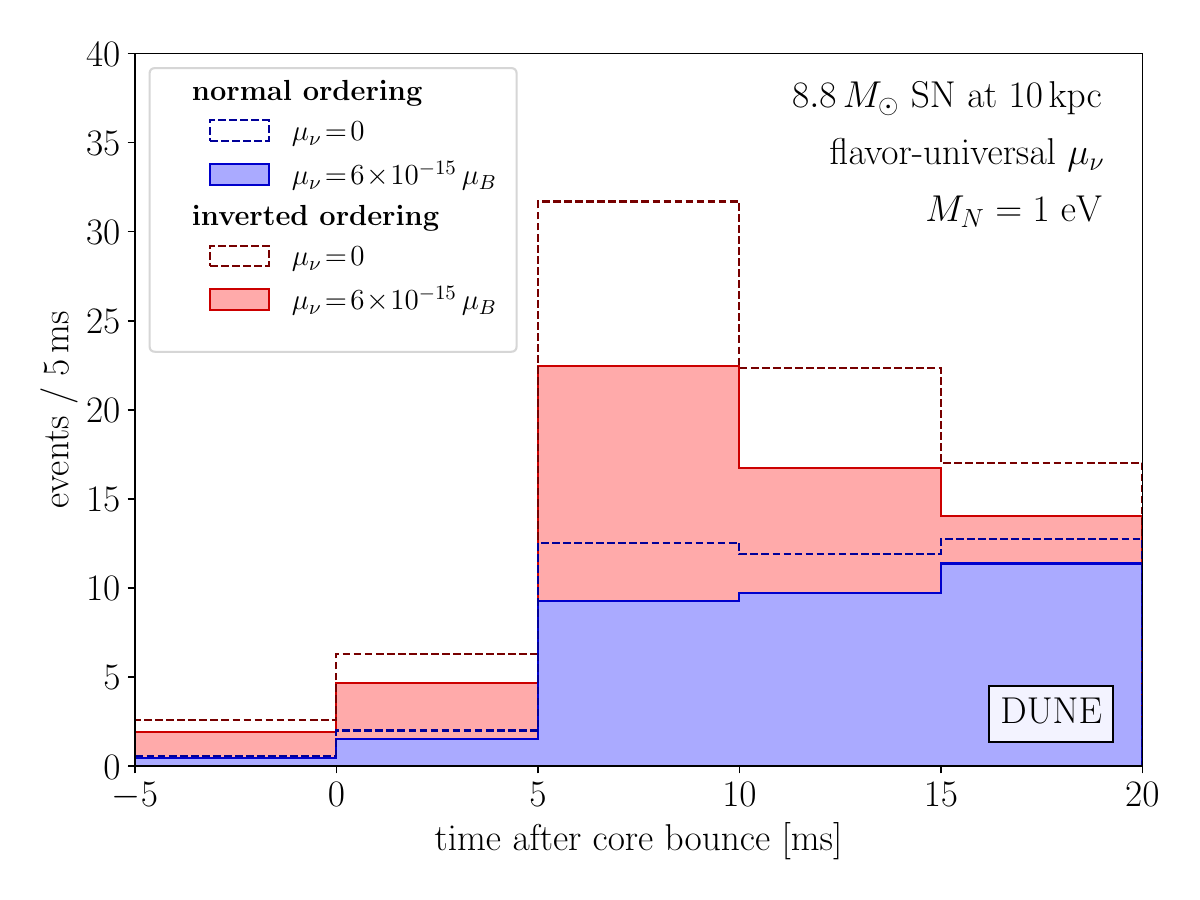}
    \end{tabular}
	\caption{Comparison of rates at Hyper-K (left) and DUNE (right) for $\mu_\nu = 0, \SI{6e-15}{\mu_B}$, $B_0 = \SI{e12}{G}$ and $M_N = \SI{1}{eV}$ for normal and inverted orderings.}
	\label{fig:rates_comparison}
\end{figure*}

It is important to note that the conversions occur between instantaneous mass eigenstates and not flavor eigenstates. Therefore, in order to obtain the correct magnetic moments to be inserted into \cref{eq:adiabaticity_parameter}, one needs to rotate the magnetic moment vector (\ref{eq:mu}) into the appropriate basis. In doing so, we find that, even when assuming all neutrino flavors to have the same magnetic moment, the magnetic moments involved in each resonance can deviate by up to an order of magnitude from one another. In particular, even though resonances RSFP-1 and 2 happen at almost the same point, the transition probabilities at the two resonances can differ significantly since the magnetic moments involved in each of them are not the same.

Assuming the MSW resonances to be adiabatic, the fluxes in the mass eigenbasis for normal hierarchy upon leaving the supernova are
\begin{subequations}
\begin{align}
	F_{\nu_3} =& P_E F_{\nu_e}^0, \\
	F_{\bar{\nu}_1} =&  P_3 F_{\bar{\nu}_e}^0, \\
	F_{\bar{\nu}_2} =& P_2 F_{\bar{\nu}'_\mu}^0 + (1 - P_3) (1 - P_2) F_{\bar{\nu}_e}^0, \\
	F_{\bar{\nu}_3} =& P_1 F_{\bar{\nu}'_\tau}^0 + (1 - P_2) (1 - P_1) F_{\bar{\nu}'_\mu}^0 \notag \\
	& + (1 - P_3) P_2 (1 - P_1) F_{\bar{\nu}_e}^0,
\end{align}
	\label{eq:fluxes_conversion_nh}%
\end{subequations}
while for inverted hierarchy, we find
\begin{subequations}
\begin{align}
	F_{\nu_2} =& P_E F_{\nu_e}^0, \\
	F_{\bar{\nu}_3} =& P_3 F_{\bar{\nu}_e}^0, \\
	F_{\bar{\nu}_1} =& P_2 F_{\bar{\nu}'_\tau}^0 + (1 - P_3) (1 - P_2) F_{\bar{\nu}_e}^0, \\
	F_{\bar{\nu}_2} =& P_1 F_{\bar{\nu}'_\mu}^0 + (1 - P_2) (1 - P_1) F_{\bar{\nu}'_\tau}^0 \notag \\
	& + (1 - P_3) P_2 (1 - P_1) F_{\bar{\nu}_e}.
\end{align}
	\label{eq:fluxes_conversion_ih}%
\end{subequations}
These equations describe the final fluxes in terms of the initial fluxes, $F_\nu^0$, after subsequent adiabatic conversions and level-crossings, depending on the transition probabilities.

\section{Sensitivity in Future Experiments}
\label{sec:exp}

The transitions described in \cref{sec:sn} would alter the neutrino fluxes arriving at Earth and would therefore leave an imprint in neutrino detectors at Earth. In the present work, we consider the signal from the neutronisation burst of a supernova in two future detectors: Hyper-Kamiokande (HK) \cite{Lagoda:2017qaj,Hyper-Kamiokande:2018ofw}, a $\SI{374}{kt}$ water Cherenkov detector, and DUNE \cite{Abi:2020evt,Abi:2020lpk}, a $\SI{40}{kt}$ liquid argon time projection chamber. HK mainly detects electron anti-neutrinos via inverse beta decay ($\bar{\nu}_e + p \to e^+ + n$), while DUNE is most sensitive to $\nu_e$ via charged current scattering on argon ($\nu_e + {}^{40} \text{Ar} \to e^- + {}^{40} \text{K}^*$). Other detection channels considered are charged current scattering on oxygen ($\nu_e (\bar{\nu}_e) + {}^{16}\text{O} \to e^\pm + X$) for HK, as well as neutral current elastic scattering on electrons ($\nu_x + e^- \to \nu_x + e^-$) for both. The cross-sections for these processes have been computed in refs. \cite{GilBotella:2003sz,Strumia:2003zx,Nakazato:2018xkv,Bahcall:1995mm,Ricciardi:2022pru}. Even though the neutrino flux from a neutronization burst is dominated by electron neutrinos, and therefore the flux at DUNE is mostly affected by RSFP-E, since HK is much more sensitive to electron anti-neutrinos than neutrinos, we find that the conversions of the antineutrinos have a sizeable impact on the detection rate at HK.

\begin{figure*}
    \centering
	\includegraphics[width= 0.8\textwidth]{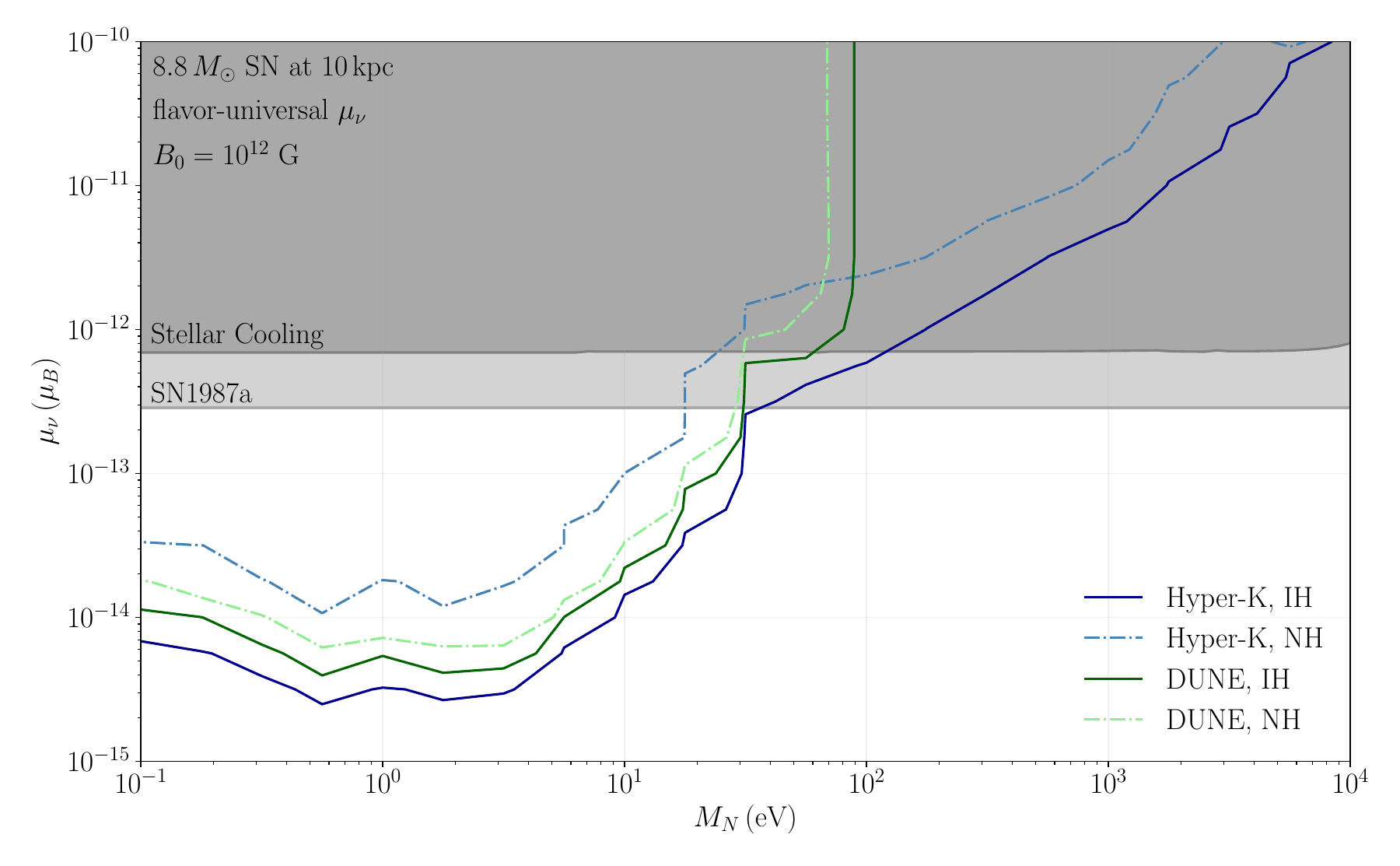}
	\caption{Projected sensitivities at $90 \%$ confidence level. Limits for Hyper-Kamiokande are shown in shades of blue, while the ones for DUNE are in shades of green. We assume flavor-universal couplings, a distance of $\SI{10}{kpc}$ to the supernova, and set $B_0 = \SI{e12}{G}$ at the iron surface of the progenitor star. Constraints from stellar cooling as well as from SN1987a (from \cite{Magill:2018jla} and \cite{Capozzi:2020cbu} respectively) are shown in grey.}
	\label{fig:exclusion}
\end{figure*}

The fluxes of neutrinos of flavor $\alpha$ arriving at Earth are then given by
\begin{equation}
	F_{\nu_\alpha} (E_\nu) = \sum_i | U_{\alpha i}|^2 F_{\nu_i} (E_\nu), 
	\label{eq:final_flux}
\end{equation}
where $F_{\nu_i}$ is the flux of neutrinos in the mass eigenstate $i$ after conversion inside the supernova, following \cref{eq:fluxes_conversion_nh,eq:fluxes_conversion_ih}.

We split the signal into five equidistant time bins from $t = - \SI{5}{ms}$ to $\SI{20}{ms}$ with respect to the core bounce and perform a $\chi^2$ analysis comparing the hypotheses of zero versus finite neutrino magnetic moments. To do this, we define
\begin{equation}
	\chi^2 = \underset{a}{\text{min}} \sum_i \frac{n_i (\mu_\nu, M_N) - (1 + a) n_i (0)}{(1 + a) n_i (0)} + \frac{a^2}{\sigma_a^2},
    \label{eq:chi-square}
\end{equation}
where $n_i (\mu_\nu, M_N)$ is the expected number of events in bin $i$ for a magnetic moment $\mu_\nu$, while $n_i (0)$ is the number events for vanishing magnetic moment. We further introduce a nuisance parameter $a$ to parametrize the uncertainty in the normalization of the initial neutrino flux. We estimate the uncertainty as $\sigma_a =0.1$. Values of $\mu_\nu$ and $M_N$ for which $\chi^2$ is larger than $4.605$ can be excluded at the $90 \%$ C.L. A comparison of the expected number of events both with and without a magnetic moment for $M_N = \SI{1}{eV}$ and $\mu_\nu = \SI{6e-15}{\mu_B}$ is shown in \cref{fig:rates_comparison}. As can be seen from the plot, the adiabatic conversions mostly affect the total normalization of the flux rather than time variations of the signal. While we did include the nuisance parameter $a$ to account for the uncertainty in the flux normalization, since we estimate its uncertainty as being rather small, at $O(10\%)$, and the magnetic moment induces a much larger reduction on the flux for some parameters, we still find a rather good sensitivity for $\mu_\nu$ in our analysis.

\section{Discussion and Conclusion}
\label{sec:cn}

In the present work, we have revisited the impact of resonant magnetic transitions of active neutrinos into a heavy sterile state on supernova measurements. We have found that this method is sensitive to magnetic moments down to a few $\SI{e-15}{\mu_B}$ for a sterile neutrino mass in the $O(\SI{}{eV})$ range. The contours at $90 \%$ confidence level assuming $B_0 = \SI{e12}{G}$ at the iron surface for a supernova at $\SI{10}{kpc}$ are summarized in \cref{fig:exclusion}. In principle, the constraints extend to $M_N$ below $\SI{e-1}{eV}$, but, in this region, $M_N$ and the active neutrino masses become comparable in size, and a more precise knowledge of the absolute neutrino masses as well as a careful treatment of the level-crossing scheme becomes necessary, which goes beyond the scope of this work.

Also shown in \cref{fig:exclusion} are constraints stemming from stellar cooling \cite{Capozzi:2020cbu} as well as from SN1987a \cite{Magill:2018jla}. The idea behind them is that sterile neutrinos can be produced via the magnetic moment interaction from the hot plasma inside the stars and escape the star unhindered, increasing its cooling rate and altering its evolution.

A similar analysis to the one presented in this paper using SN1987a data was attempted by applying the same methods to the time-integrated flux. Since, however, the main effect of resonant magnetic conversion on supernova neutrino measurements is an overall deficit of the flux, and the total luminosity is a free parameter of the fit, we were unable to derive new constraints from it. We have not analyzed the time-dependent flux, but we don't expect it to yield much better results.

We find the strongest constraints for $M_N$ at around $\SI{1}{eV}$. In this case, for neutrinos with a typical energy $p = \SI{10}{MeV}$, the resonances RSFP-1, 2 and $E$ happen at $r \approx r_0$, where the magnetic field is close to its maximum, and $d V/d r$ has already decreased considerably, as mentioned in \cref{sec:sn}. We also see from \cref{fig:exclusion} that the sensitivity for inverted hierarchy is better than for normal hierarchy in both detectors. This can be explained by the larger flux of electron neutrinos expected in the case of inverted hierarchy, since electron neutrinos produced inside the supernova leave it as $\nu_2$ mass eigenstates, as opposed to $\nu_3$ in inverted ordering. Since $\nu_2$ has a much larger electron component than $\nu_3$, the electron neutrino flux at Earth is expected to be significantly enhanced in inverted hierarchy compared to the normal one. However, we also find the difference to be more pronounced for HK and DUNE. This can be understood by looking again at \cref{eq:fluxes_conversion_nh,eq:fluxes_conversion_ih}.

As we discussed in \cref{sec:sn}, RSFP-3 happens deep inside the supernova, where $d V_\nu/dr$ is large and therefore $\gamma_\text{res}$ is small, so that we can expect $P_3 \approx 1$. This means that the largest effect on the electron anti-neutrino flux comes from RSFP-1 and 2. In inverted hierarchy, these resonances impact both $\bar{\nu}_1$ and $\bar{\nu}_2$, which both have large electron components, while in normal hierarchy they affect $\bar{\nu}_2$ and $\bar{\nu}_3$, leaving $\bar{\nu}_1$, with the largest electron component of all, unchanged. Since HK is mostly sensitive to electron anti-neutrinos, this produces an additional source of discrepancy between the two hierarchies.

It should be noted, however, that our constraints depend on the precise magnetic field of the supernova. Since the adiabaticity parameter depends on the product $\mu_\nu B$, the sensitivity in $\mu_\nu$ is inversely proportional to the magnetic field strength. In addition to this, while the first derivative of the geometric phase $\dot{\phi}$ only affects the resonance point and therefore shifts our curves in the $M_N$ axis, a large turbulent component to the magnetic field may compromise adiabatic conversion altogether if the second derivative of the geometric phase $\ddot{\phi}$ is much larger than $d V/d r$ \cite{Smirnov:1991ia}.

While modeling the magnetic field of supernovae is still challenging, this field has been rapidly evolving, with studies investigating the emission of the SN1987a remnant to infer properties of its progenitor \cite{Orlando:2015dfa,Orlando:2018lbj,Petruk:2022urq}. We believe that, by the time the next galactic supernova occurs, we will be able to better understand its magnetic field profile and, from this, derive concrete constraints on (or discover) large neutrino magnetic moments.

\hspace{1cm}

\section*{Acknowledgements}

I would like to thank Joachim Kopp for his invaluable support and for proofreading the draft. This work is supported by
the Collaborative Research Center SFB 1258 of the German Research Foundation (DFG).

\bibliography{references.bib}

\end{document}